\documentclass[showpacs,amsmath,amssymb,floatfix,aps,pre,letterpaper,twocolumn]{revtex4} 
\usepackage[dvips]{graphicx} 
\usepackage[utf8]{inputenc}
\usepackage{color,rotating}
\usepackage{amsmath} 
\usepackage[english]{babel}
\usepackage{natbib}
\newcommand{\matris}[1]{\mathbf{#1}}
\newcommand{\kBT}{k_\text{B}T}
\begin{document}
\title{Decay times in turnover statistics of single enzymes}
\date{July 10, 2008}  \author{Martin Lind\'en} \email{linden@kth.se}
\affiliation{Department of Theoretical Physics, Royal Institute of
  Technology, SE-10691 Stockholm, Sweden}
\pacs{87.10.-e,82.39.-k,02.50.-r,05.40.-a}
\begin{abstract}
The first passage times for enzymatic turnovers in non-equilibrium
steady state display a statistical symmetry property related to
non-equilibrium fluctuation theorems, that makes it possible to
extract the chemical driving force from single molecule trajectories
in non-equilibrium steady state. Below, we show that the number of
decay constants needed to describe the first passage time distribution
of this system is not equal to the number of states in the first
passage problem, as one would generally expect.
Instead, the structure of the kinetic mechanism makes half of the
decay times vanish identically from the turnover time
distribution. The terms that cancel out correspond to the eigenvalues
of a certain sub-matrix of the master equation matrix for the first
exit time problem. We discuss how these results make modeling and data
analysis easier for such systems, and how the turnovers can be
measured.
\end{abstract}
\maketitle Enzymes are vital to most biochemical reactions, to
increase reaction speed and as active components in cellular
regulatory networks. Observations of the fluctuations on the single
molecule level can lead to new insights into enzymatic mechanisms, by
revealing more detailed information than ensemble averages measured in
bulk
experiments\cite{shaevitz05,kolomeisky05,wang06,qian06,min05,tsygankov07a,linden07}.
The rapid development in single molecule techniques has made it
possible to directly observe turnover events of single enzymes in many
systems (see e.g., Ref.~\citep{moerner07} and refs. therein).  This
motivates continuing theoretical interest in stochastic kinetics.  For
example, it was recently shown how non-equilibrium fluctuation
theorems make it possible to extract the chemical driving force from
turnover traces of single enzymes\cite{qian06,min05}.

Here we consider the statistical properties of reversible enzymatic
turnovers, and derive another useful property of the turnover
times. Their distribution is a sum of exponentially decaying terms,
and the number of terms is usually expected to reflect the number of
states in the underlying first passage problem.  We show that the
number of terms in the actual distributions are only half of the
expected number, due to the periodicity of the problem. Moreover, we
discuss an earlier suggestion\citep{qian06} of how to detect turnover
events, and conclude that it does not correspond to the first passage
problem for turnover times. We have previously addressed an analogous
issue for stepping motor proteins, and shown that it can lead to
systematic misinterpretations of experimental
data\cite{tsygankov07a,linden07}.  Our results point to modifications
in previously suggested experiments, and also simplify theoretical
analysis of turnover time distributions.

In the next section, we introduce our model and the results. After
that, we discuss how turnover times can be detected in reversible
single molecule experiments. We then derive our main result, and
finally discuss some implications.

\begin{figure}
  \includegraphics[width=8.7cm]{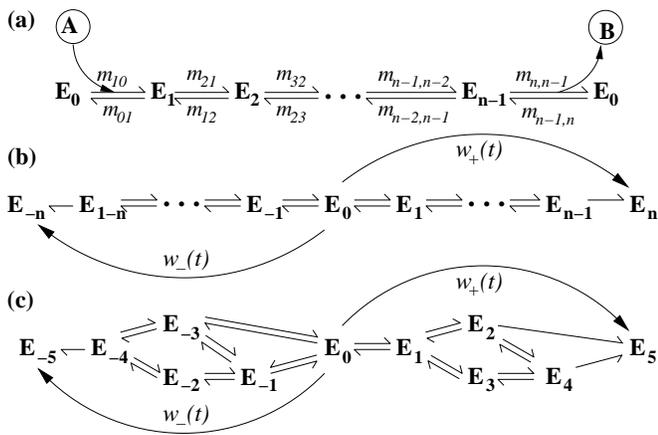}
  \caption{(a) A multi-step enzyme reaction converting A to B. The
    turnover time distributions $w_\pm(t)$ are given by the first exit
    time problem in (b). The enzyme starts in state E$_0$ and is
    absorbed in E$_n$ or E$_{-n}$. The rate constants $m_{10}$ and
    $m_{n-1,n}$ are pseudo-first-order, i.e., proportional to the
    concentrations of A and B respectively.  In the more general
    scheme studied in Ref.~\citep{wang06}, arbitrary transitions
    within a cycle are allowed. An example of this, with $n=5$
    different enzymatic states, is illustrated in (c). By periodicity,
    state E$_k$ is equivalent to E$_{k+n}$, and
    $m_{ij}=m_{i+n,j+n}$.}\label{path}
\end{figure}

\paragraph*{Model.} 
Following \citet{qian06}, we start with a simple sequential kinetic
model of an enzyme reaction, sketched in Fig.~\ref{path}(a), where a
substrate A is converted to a product B through several intermediate
states. The overall concentrations of substrate and product molecules
are assumed to be kept constant, so that a non-equilibrium steady
state is maintained.

As sketched in Fig.~\ref{path}(b), a $\pm$ turnover is defined as the
first arrival in one of the empty states $\text{E}_{\pm n}$, after
start in state $\text{E}_\text{0}$. If a turnover is completed at
$t=0$, the integrated turnover times $w_\pm(t)$ is the probability
that the next turnover is a $\pm$, and occurs at time $t$ or
earlier. Thus, the turnovers are equivalent to the cycle completion
events associated with the work of \citet{hill89}.
Microscopic reversibility leads to a symmetry property for the forward
($+$) and backward ($-$) turnover times\cite{derenyi99}, namely,
$w_+(t)=e^{\Delta \mu/\kBT}w_-(t)$ \cite{qian06,wang06}. Here,
$\frac{\Delta \mu}{\kBT}=\ln\frac{m_{10}m_{21}\ldots
  m_{n,n-1}}{m_{01}m_{12}\ldots m_{n-1,n}}$ is the chemical driving
force, and $m_{ij}$ is the rate of the transition $\text{E}_j\to
\text{E}_i$.

The turnover time distributions are of the general form
\begin{equation}
  w_\pm(t)=\alpha^\pm_0+\sum_{k=1}^N\alpha^\pm_ke^{\lambda_kt}.
\end{equation}
The characteristic decay times $\tau_k=-1/\lambda_k$ and prefactors
$\alpha_k^\pm$ depend on the transition rates and topology of the
underlying kinetic mechanism. Hence, this mechanism can be studied by
fitting theoretically predicted distributions to experimental data.

The underlying first passage problem is governed by a system of linear
master equations \cite{vankampen}, one equation for each state from
which the systems escapes.  Since one generally expects a matrix of
dimension $N$ to have $N$ eigenvalues, a simple and common way to
estimate the number of states is to count how many exponential terms
are needed to fit the first passage time distribution.  As illustrated
in Fig.~\ref{path}, the turnover events correspond to escape events
from states E$_{1-n}$, E$_{2-n}$,\ldots,E$_{n-1}$. Hence, the number
of states in the first passage problem is $N=2n-1$ in this case, where
$n$ is the number of intermediate states of the enzyme-substrate
complex. As shown below, the structure of this first passage problem,
as well as the more general one studied by \citet{wang06}, makes $n-1$
of the coefficients $\alpha_k^\pm$ in $w_\pm(t)$ vanish.  This leaves
only $n$ terms in the distribution, i.e., the above estimate fails by
a factor two. Before we derive this, we discuss how turnover times
can be detected in reversible single molecule experiments.
\paragraph*{Detecting enzymatic turnovers.}
Single enzyme experiments using fluorescence techniques often probe
the state of a enzyme-substrate
complex\cite{english06,lu98,kou05,min06}, but do not report directly
on the number of turnovers. In our example model, a realistic
possibility is that the empty states (E$_0$, E$_{\pm n}$, \ldots) can
be experimentally distinguished from the other states, but not from
each other. If the product concentration is kept very low, it is safe
to assume that each departure from an empty state starts a new forward
turnover. However, detecting individual forward and backward turnovers
in conditions where backward turnovers are possible is more
complicated, as the following discussion will show.

In their proposal to measure $\Delta\mu$ directly from turnover
traces, \citet{qian06} suggested that individual turnover times could
be measured by monitoring the net number $\nu_B(t)$ of product
molecules, as they are released and absorbed by the reaction
\mbox{$\text{E}_{n-1}\rightleftharpoons\text{E}_n$}. However, this is
equivalent to monitoring the position of a processive motor protein.
A closer examination reveals that this situation corresponds to a
different first passage problem, with quite different statistical
properties. This discrepancy can lead to large systematic errors in
the estimate for $\Delta\mu$\cite{tsygankov07a,linden07}.

To see why that measurement will not detect turnovers, note that the
turnover event starts in state E$_0$, and finishes when an enzymatic
cycle is completed, i.e., when either E$_n$ or E$_{-n}$ is reached for
the first time \cite{qian06,wang06,kolomeisky05}. However, $\nu_B(t)$
does not change during the reaction
E$_{1-n}\to\text{E}_{-n}$. Therefore, backward turnovers cannot be
detected by only monitoring changes in $\nu_B$.

The attractive statistical properties of turnover
times\cite{qian06,wang06,min05} motivate a consideration of how they
could be measured, using slightly different experimental setups.  One
possibility would be to monitor both substrate and product molecules,
but one could also imagine various setups involving fluorescence
techniques with multiple fluorescence levels, in the spirit of the
experiment that demonstrated bi-directional rotation in ATP
synthase\cite{diez04}.  To summarize, it is important to make sure
that the theoretical first passage problem describes the actual
experimental situation.
\paragraph*{Number of decay times in $w_\pm(t)$.}
We now derive our main result, i.e., that the number of exponential
terms in the turnover time distributions $w_\pm(t)$ are not given by
the number of states in the first passage problem, $2n-1$, as one
might expect\cite{qian06}. Instead, $w_\pm(t)$ only contains $n$
terms.  The decay constants that drop out are the eigenvalues of a
certain sub-matrix of the master equation matrix for the first exit
time problem. This result may simplify practical calculations
considerably.

The turnover times studied by \citet{qian06} are the solutions of the
first exit problem illustrated in Fig.~\ref{path}(b).  If we label the
states $-n,-n+1,\ldots,n-1,n$, the system starts in state 0 and is
absorbed in states $\pm n$. Let $q_k(t)$ be the probability of being
in state $k$ a time $t$ after starting in state $0$. The $q_k(t)$ are
governed by the master equation
\begin{equation}\label{e2}
  \partial_t q_i(t)=\sum_{j\ne i}\Big(m_{ij}q_j(t)-m_{ji}q_i(t)\Big),
  \quad -n<i<n,
\end{equation}
with initial condition $q_j(0)=\delta_{j,0}$.  Since $\pm n$ are
absorbing states, the integrated turnover time distribution functions
are given by
\begin{equation}\label{e3}
w_\pm(t)=q_{\pm n}(t)=\!\int_0^t\!\frac{d q_{\pm n}(t)}{dt}dt=
\!\sum_{j=1-n}^{n-1}m_{\pm n,j}\!\int_0^t\!  q_j(t)dt.
\end{equation}
Note that $w_\pm(t)$ are not normalized to unity. Instead, the
fractions $p_\pm$ of $\pm$ turnovers are given by
$p_\pm=\lim_{t\to\infty} w_\pm(t)$.  Introducing the matrix
$\matris{M}$ and vector $\vec{q}(t)$ with elements
\begin{align}
  M_{ij}&=m_{ij}-\delta_{ij}\!\!\sum_{k=1-n}^{n-1}\!\!m_{ik}, 
  \quad -n<i,j<n,\\ 
  \vec{q}(t)&=[q_{-n+1}(t),q_{-n+2}(t),\;\ldots,\;q_{n-1}(t)]^\text{T},
\end{align}
the solution of Eq.~\eqref{e2} can be written
$\vec{q}(t)=e^{t\matris{M}}\vec{q}(0)$.

We restrict our attention to the generic situation where $\matris{M}$
can be diagonalized\cite{nondegenerate}. In this case, $\vec{q}(t)$
can be expressed in terms of right and left eigenvectors
$\vec{a}^{(k)}$ and $\vec{b}^{(k)}$ of $\matris{M}$, namely
\begin{equation}\label{e4}
  \vec{q}(t)=  \!\sum_{k=1}^{2n-1}e^{t\lambda_k}\vec{a}^{(k)}
  \big(\vec{b}^{(k)}\cdot\vec{q}(0)\big).
\end{equation}
Note that the eigenvalues $\lambda_k$ need not all be
distinct\cite{schnakenberg76}.  Among the $2n-1$ terms, we look for
left eigenvectors $\vec{b}^{(k)}$ that are orthogonal to the initial
condition $\vec{q}(0)$.  Those terms drop out of Eq.\ \eqref{e4}, and
hence from $w_\pm(t)$ as well.  We introduce
\begin{equation}
  \begin{split}
  \vec{y}_+&=[m_{1-n,0}      ,\ldots , m_{-1,0}]^T,
  \vec{y}_-=[m_{1,0}         ,\ldots , m_{n-1,0}]^T\!\!,\\
  \vec{v}_+&=[m_{0,1-n}      ,\ldots , m_{0,-1}]^T,
  \vec{v}_-=[m_{0,1}         ,\ldots , m_{0,n-1}]^T\!\!,\\
  \vec{b}_-^{(k)}&=[b_{1-n}^{(k)},b_{2-n}^{(k)},\ldots , b_{-1}^{(k)}]^T,
  \vec{b}_+=[b_1^{(k)},b_2^{(k)}         ,\ldots , b_{n-1}^{(k)}]^T\!\!,\\
  \end{split}
\end{equation}
and take $\matris{Y}$ as the $(n-1)\times(n-1)$ matrix with elements
$Y_{ij}=M_{ij}$ for $0<i,j<n$, i.e., the master equation matrix for
the first exit problem from states $1,2,\ldots,n-1$.  Using the
periodicity of the transition rates, $m_{i,j}=m_{i+n,j+n}$, and a
'bottleneck' property of state $0$, $m_{i,j-n}=m_{j-n,i}=0$ for
$0<i,j<n$, the left eigenvalue problem for $\matris{M}$ can be written
\begin{equation}\label{Mekv}
  \matris{M}^T\vec{b}^{(k)}= \left[\begin{array}{ccc}
  \matris{Y}^T&\vec{y}_-&\matris{0}\\ \vec{v}_-^T&M_{00}&\vec{v}_+^T\\
  \matris{0}&\vec{y}_+&\matris{Y}^T\\
    \end{array}
  \right]\left[\begin{array}{l}
      \vec{b}_-^{(k)}\\b_0^{(k)}\\\vec{b}_+^{(k)}
    \end{array}\right]=\lambda_k\vec{b}^{(k)},
\end{equation}
where $\matris{0}$ is the $(n-1)\times (n-1)$ zero matrix.  This
structure of $\matris{M}$ also holds for the more general turnover
time problem studied in Ref.\ \citep{wang06} and illustrated in
Fig.~\ref{path}(c). In the special case studied in
Ref.\ \citep{qian06} (Fig.~\ref{path}(b)), $\matris{M}$ and
$\matris{Y}$ are tridiagonal, in which case $\vec{y}_\pm,\vec{v}_\pm$
have only one non-zero element each.  Setting
$\vec{b}^{(k)}\cdot\vec{q}(0)=b_0^{(k)}=0$ in Eq.\ \eqref{Mekv} gives
\begin{align}
    \matris{Y}^T \vec{b}_-^{(k)} &=\lambda_k \vec{b}_-^{(k)},\\
    \vec{v}_- \cdot \vec{b}_-^{(k)}+\vec{v}_+ \cdot
    \vec{b}_+^{(k)}&=0,\\ \matris{Y}^T \vec{b}_+^{(k)} &=\lambda_k
    \vec{b}_+^{(k)}.
\end{align}
The solutions are given by the eigenvalues of $\matris{Y}^T$, which
are also eigenvalues of $\matris{M}$.  Since these have equal
algebraic and geometric multiplicity by assumption, there are $n-1$
solutions, corresponding to terms that do not contribute to the
turnover time distributions in Eq.~\eqref{e4}. Hence, $w_\pm(t)$
contains at most $n$ exponential terms: those where $\lambda_k$ is an
eigenvalue of $\matris{M}$, but not of $\matris{Y}$.
\paragraph*{Turnover time distribution for sequential models.}
Analytical expressions for the turnover time distributions is useful
for efficient parameter extraction. Our result for the number of
exponential terms in $w_\pm(t)$ makes the derivation of such
expressions easier, and extends the range of system sizes that can be
treated analytically.

To illustrate this, we analyze the sequential model in
Fig.~\ref{path}(a), using the ansatz
\begin{equation}\label{ansatz}
  w_\pm(t)=p_\pm(1+\alpha_1 e^{\lambda_1t}+\ldots
+\alpha_{n} e^{\lambda_{n}t}),
\end{equation}
with $\alpha_k=\alpha_k^\pm/p_\pm$, together with the initial
conditions for $w_\pm(t)$. As shown by \citet{qian06} (their
Ref. [24]), the sequential models satisfy
$p_\pm^{-1}w_\pm(0)=1+\sum_k\alpha_k=0$, and $p_\pm^{-1}\partial_t^m
w_\pm(0)=\sum_k\lambda_k^{m}\alpha_k=0$ for $1\le m\le n-1$.
This leads to a Vandermonde type system of equations,
\begin{equation}\label{vandermonde}
  \left[\begin{array}{cccc} 1&1&\ldots&1\\
    \lambda_{1}&\lambda_{2}&\ldots&\lambda_{n}\\
    \vdots&\vdots &    \ddots&\vdots\\
    \lambda_{1}^{n-1}&\lambda_{2}^{n-1}&\ldots&
    \lambda_{n}^{n-1}\\
  \end{array}\right]
  \left[\begin{array}{c}\alpha_1\\\alpha_2\\
    \vdots\\\alpha_n\end{array}\right]
    =
  \left[\begin{array}{c}-1\\0\\
\vdots\\0\end{array}\right].
\end{equation}
Solving with Cramer's rule for the normalized distribution
$p_\pm^{-1}w_\pm(t)$, we get
\begin{equation}\label{seqdist}
p_\pm^{-1}w_\pm(t)=1+(-1)^n\sum_{k=1}^{n}e^{\lambda_kt}\prod_{m\ne k}
\Big( \frac{\lambda_m}{\lambda_k-\lambda_m}\Big).
\end{equation}
This is the distribution of a sum $N$ of independent exponential
random variables with mean values $|\lambda_1|^{-1},
|\lambda_2|^{-1},\ldots, |\lambda_n|^{-1}$.  
The reduced number of unknown coefficients $\alpha_k$ simplifies the
analytical computation significantly, especially so for sequential
models, where $\sum_k\lambda_k^m\alpha_k\ne 0$ for \mbox{$m\ge n$}
\cite{qian06}.

Analytical calculation of the eigenvalues $\lambda_k$ means finding
the roots of a characteristic polynomial.  (For large systems, the
eigenvalues must be found numerically. In these cases, root-finding in
characteristic polynomial is usually not the best
method\cite{numres}.) Since the \mbox{$n-1$} non-contributing time
constants are the eigenvalues of $\matris{Y}$, these can be removed
from the eigenvalue equation in advance, hence reducing the problem
from root-finding in the characteristic polynomial of $\matris{M}$,
$P_\matris{M}(\lambda)=\det(\matris{M}-\lambda\matris{I})$, which has
degree $2n-1$, to root-finding in the polynomial
$P_\matris{M}(\lambda)/P_\matris{Y}(\lambda)$, which has degree $n$.
This makes it feasible to compute decay constants analytically for
larger systems.
\paragraph*{Conclusion}
We have demonstrated that enzymatic turnover times constitute a
counterexample to the expectation that the number of states in a first
passage time problem is equal to the number of exponential terms in
the first passage time distribution. Instead, the number of terms is
in this case equal to the number of states per cycle.
This number is an important characteristic of the kinetic mechanism of
an enzyme, and our results make it possible to estimate it correctly
from time series of turnover times.

Furthermore, our results make it easier to derive (semi)analytical
expressions for the turnover time distributions, thus simplifying
modeling and data analysis. The approach demonstrated above for a
sequential model also works for large systems, if the eigenvalues are
computed numerically.

Finally, we have supplied an important clarification to an earlier
suggestion\cite{qian06} on how to detect turnover events. This should
make our results, together with earlier
predictions\cite{min05,qian06,wang06}, into useful analysis tools for
future experiments, and a further reason to study enzymatic turnovers
under reversible conditions.

\begin{acknowledgments}
The author is grateful for constructive input from Hong Qian, Mats
Wallin, Michael E.~Fisher, and Denis Tsygankov, and financial support
from the Royal Institute of Technology and the Wallenberg Foundation.
\end{acknowledgments}


\begin{thebibliography}{20}
\bibitem[{non()}]{nondegenerate}
\bibinfo{note}{This is the generic situation in the same sense as for square
  matrices in general, i.e., the set of $\matris{M}$-matrices contains a dense
  subset of matrices with non-degenerate eigenvalues. This follows from
  corollary 1 of Ref.~\cite{hartfiel95}.}
\end{thebibliography}

\end{document}